\documentstyle[aps,preprint]{revtex}
\begin{document}
\title{Finite Temperature Path Integral Method for Fermions and 
Bosons: a Grand Canonical Approach} 
\author{M. Skorobogatiy, J. Joannopoulos}
\address{Department of Physics, MIT, Cambridge, USA}

\date{\today}
\maketitle

\smallskip

\begin{abstract}
The calculation of the density matrix for fermions and bosons in the Grand 
Canonical 
Ensemble allows an efficient way for the inclusion of fermionic and 
bosonic statistics at all temperatures. It is shown that in a Path Integral 
Formulation fermionic density matrix can be expressed via an 
integration over a novel representation of the universal temperature 
dependent functional. While 
several representations for the universal functional have already 
been developed, 
they are usually presented in a form inconvenient for computer 
calculations. In this work we discuss a new representation for 
the universal functional in terms of Hankel functions which is 
advantageous for computational applications. Temperature scaling
for the universal functional and its derivatives are also introduced thus 
allowing an efficient rescaling rather then recalculation of the 
functional at different temperatures. A simple illustration of the method of 
calculation of density profiles in Grand Canonical ensemble is presented 
using a system of noninteracting electrons in a finite confining potential.  
\end{abstract}

\newpage
\narrowtext

\section{Introduction}
The standard treatment of fermionic and bosonic systems in canonical ensemble 
using Path Integral Formalism is based on the isomorphism between a 
system of interacting quantum particles and a set of interacting 
classical polymer rings \cite{Chandler}. For a system of bosons all the 
ring contributions to the density matrix come with the same sign thus 
allowing an effective Monte-Carlo sampling of the relevant terms 
\cite{Ceperley1}. However, for the case of fermions situation is more 
complicated as different ring contributions to the density matrix come with 
alternating signs thus making statistical fluctuations so big that an 
unreasonably long averages are required to provide for the meaningful
results. While several promising attempts were made to circumvent the
sign problem \cite{Kalos,Ceperley2} there is still much to be done before 
this problem will be solved.  
 
However, instead of operating in a canonical ensemble for fermionic and 
bosonic systems it might be more advantageous to work in the Grand Canonical 
Ensemble. To see why, as soon as one knows the normal excitations $E_i$ of 
the system the grand canonical partition function is
 then\begin{equation}
\Xi (\beta)=\pm \sum_{i=1}^{\infty} log(1 \pm Zexp(-\beta E_i))
\end{equation}
where $Z=exp(\beta \mu)$. The density matrix can then be expressed as
\begin{equation}
\rho(r,r';\beta,\mu)=<r|\frac{1}{Z^{-1}exp(\beta H) \pm 1}|r'>
\end{equation}
where $H$ is the Hamiltonian. By reformulating the problem 
in this fashion, one need not worry about the explicit 
antisymmetrization or symmetrization of $|r>$ and $|r'>$
states. The problem of finding $\Xi (\beta)$ and $\rho(r,r';\beta,\mu)$ is
re-expressed in terms of the normal excitations of a 
system. Thus if in the canonical formulation we had $N$ interacting 
fermionic or bosonic particles with a density matrix $exp(-\beta H_0)$ 
and a properly symmetrized wave function, in the grand canonical formulation 
we can in principle reduce the problem to a system of non interacting 
particles described by the density matrix $\frac{1}{Z^{-1}exp(\beta H) 
\pm 1}$. This appears much simpler than the problem in a canonical 
formulation but there is one problem. The question is how 
to find a spectrum of normal excitations. For fermionic systems the 
problem is solved in principle in the Kohn Sham (KS) density functional 
formalism \cite{Kohn}. According to (KS) the problem of $N$ interacting 
fermions with a Hamiltonian $H_0$ can be mapped into a problem of $N$ 
noninteracting particles described by a modified Hamiltonian $H$ which 
is, in turn, the functional of the density (see the review \cite{Payne})
\begin{equation}
H=-\frac{\hbar^2}{2m}\bigtriangledown^2+V_{ion}(r)+V_{H}(r)+V_{XC}(r)
\end{equation}     
where $V_{ion}(r)$ is an electron-ion interaction potential,
\begin{equation}
V_{H}(r)=e^2 \int \frac{\rho(r',r';\beta)}{|r-r'|} d^3 r'
\end{equation} 
is the Hartree potential of the electrons and finally, 
\begin{equation}
V_{XC}(r)=\frac{\delta E_{XC}(\rho(r,r;\beta))}{\delta \rho(r,r;\beta)}
\end{equation}
is a universal exchange correlation electron density functional.

Thus, a self-consistent algorithm for finding the density matrix in the 
Grand Canonical Formulation would be implemented as following\\
a) Use $\rho(r,r;\beta)$ to find a new effective Hamiltonian $H$\\
b) Given $H$ calculate a new $\rho(r,r;\beta)$ from
\begin{equation}
\rho(r,r';\beta,\mu)=<r|\frac{1}{Z^{-1}exp(\beta H)\pm 1}|r'>
\end{equation}\\
c) Update the chemical potential $\mu$ (which comes via $Z=exp(\beta 
\mu)$)\\ 
d) If self-consistency is not achieved go to a)

Up to this point the algorithm is rather general and can be implemented in 
many ways. One of the implementations of this algorithm for fermions is 
originally due to 
S. Goedecker \cite{Goedecker} and lately extended by Baer \cite{Baer}.
The most important step is to implement part b). So that 
\begin{equation}
\rho(r,r';\beta,\mu)=<r|\frac{1}{Z^{-1}exp(\beta H)+1}|r'>=
\sum_{k=1}^{\infty} \sum_{k'=1}^{\infty} <r|k><k|\frac{1}
{Z^{-1}exp(\beta H)+1}|k'><k'|r'> 
\end{equation}
where $<r|k>$ is any convenient basis set. By approximating
$\frac{1}{Z^{-1}exp(\beta H)+1}=Pol(H;\beta)$ using Chebyshev 
Polynomials Goedecker et. al obtained an efficient way of calculating the 
density matrix using $Pol(H;\beta)$ as a propagator.

An alternative and potentially more beneficial method for implementation of 
step b) is to use the Path Integral approach \cite{Wang,Alavi}. 
In the Path Integral Formalism to evaluate the density matrix one would 
insert $P$ auxiliary states thus reducing the problem to the evaluation 
of $<r|(\frac{1}{exp(\beta(H-\mu))\pm 1})^{\frac{1}{P}}|r'>$. At high 
temperature 
\begin{eqnarray}
<r|(\frac{1}{exp(\beta (H-\mu))\pm 1})^{\frac{1}{P}}|r'> \sim
<r|exp(-{\frac{\beta}{P}} (H-\mu))|r'> \sim \\ \nonumber 
<r|exp(-\frac{\beta}{P} T_{kin})|r'>exp(-\frac{\beta}{P} V(r'))
\end{eqnarray}        
so the conventional use of the Trotter approximation for the splitting of the 
Hamiltonian into kinetic and potential energy is applicable.
At low temperatures $\beta \rightarrow 0$ in the case of fermion 
statistics
\begin{equation}
<r|(\frac{1}{exp(\beta (H-\mu))})^{\frac{1}{P}\pm 1}|r'> \sim \theta(\mu-H)
\end{equation}
where $\theta(x)$ is a step function and the propagator becomes invariant 
with respect to $P$.
So the major difference between using the grand canonical approach in the 
Path Integral Formulation is the approximation of the intermediate 
propagator. As a function of $P$, propagator 
$<r|(\frac{1}{exp(\beta (H-\mu))\pm 1})^{\frac{1}{P}}|r'>$
is very easy to estimate 
at high temperature by factoring out the kinetic and potential term. At low
temperatures the propagator becomes $P$-independent thus substantially 
hindering its use in calculations and thus new methods need to be 
implemented for the evaluation of this propagator. In this work we 
present a tractable solution valid at all temperatures.

\section{All Temperature Solution}
The problem of evaluation of the fermionic density matrix at zero temperature 
allows an exact solution in the Path Integral Formulation in terms of Bessel 
functions \cite{Harris} . For the case of non-zero temperatures the
problem has also been solved with the fermionic density matrix 
represented in integral form \cite{Yang}.  

Here we will show that there exists a more computationally convenient 
representation for the fermionic and bosonic density matrix in terms of the 
Hankel functions of the first kind.

To calculate the density matrix in the Grand Canonical Ensemble we will 
exploit the property of a meromorphic function been equal to a summation 
over its poles and residues at those poles. 
\begin{equation}
\frac{1}{exp(\beta(H-\mu))\pm 1}=\pm\frac{1}{2}\pm\sum_{n=-\infty}^{+\infty}
\frac{1}{iw_n-\beta(H-\mu)}
\end{equation}
where $w_n=\pi(2n+1)$ for fermions and $w_n=2\pi n$ for bosons. The density 
matrix now can be written as 
\begin{equation}
<r|\frac{1}{exp(\beta(H-\mu))\pm 1}|r'>=\pm\frac{\delta(r-r')}{2}\pm
\sum_{n=-\infty}^{+\infty} <r|\frac{1}{iw_n-\beta(H-\mu)}|r'>
\label{eq0}
\end{equation}

In this form it is still difficult to make a transition to a Path Integral 
Formalism. We notice, however, that for $w_n>0$
\begin{equation}
\frac{1}{iw_n-\beta(H-\mu)}=-i\int_{0}^{+\infty}dt exp(-tw_n-it\beta(H-\mu))
\label{eq1}
\end{equation}
and for $-w_n$
\begin{equation}
\frac{1}{-iw_n-\beta(H-\mu)}=i\int_{0}^{+\infty}dt exp(-tw_n+it\beta(H-\mu))
\label{eq2}
\end{equation}
Thus, evaluation of $<r|\frac{1}{iw_n-\beta(H-\mu)}|r'>$ is now substantially
simplified and we can change to a Path Integral Formulation.
\begin{equation}
<r|\frac{1}{iw_n-\beta(H-\mu)}|r'>=-i\int_{0}^{+\infty}dt 
exp(-tw_n)<r|exp(-it\beta(H-\mu))|r'>
\end{equation}
Element $<r|exp(-it\beta(H-\mu))|r'>$ is nothing else but a real time
propagator. Its evaluation in a Path Integral Formalism is trivial and 
leads to
\begin{eqnarray}
<r|exp(-it\beta(H-\mu))|r'>=lim_{t-i\eta \rightarrow t} 
(\frac{-imP}{2\pi \hbar t\beta})^{\frac{3P}{2}} 
exp(i(\frac{ml_p^2}{2\hbar t}+\frac{\hbar k_p^2t}{2m}))
\end{eqnarray}
where
\begin{eqnarray}
\frac{\hbar^2 k_p^2}{2m}=\beta(\mu-\frac{1}{P}
(\frac{V({\bf r})+V({\bf r'})}{2}+\sum_{i=1}^{P-1} V({\bf r^{(i)}})))
\end{eqnarray}  
\begin{eqnarray}
l_p^2=\frac{P}{\beta}(({\bf r}-{\bf r^{(1)}})^2+\sum_{i=1}^{P-2}
({\bf r^{(i)}}-{\bf r^{(i+1)}})^2+({\bf r^{(P-1)}}-{\bf r'})^2)
\end{eqnarray}
Finally, for $w_n>0$
\begin{eqnarray}
<r|\frac{1}{iw_n-\beta(H-\mu)}|r'>=-i\int_{0-i\eta}^{+\infty-i\eta}dt
exp(-tw_n) (\frac{-imP}{2\pi \hbar t\beta})^{\frac{3P}{2}}
exp(i(\frac{ml_p^2}{2\hbar t}+\frac{\hbar k_p^2t}{2m}))
\end{eqnarray}
and for $-w_n$
\begin{eqnarray}
<r|\frac{1}{-iw_n-\beta(H-\mu)}|r'>=i\int_{0+i\eta}^{+\infty+i\eta}dt
exp(-tw_n) (\frac{imP}{2\pi \hbar t\beta})^{\frac{3P}{2}}
exp(-i(\frac{ml_p^2}{2\hbar t}+\frac{\hbar k_p^2t}{2m}))
\end{eqnarray}

Using the integral representation of the Hankel's functions \cite{Gradshteyn} 
\begin{equation}
H^{(1)}_n(z)=-\frac{i}{\pi}exp(-\frac{1}{2}in\pi)z^n \int_0^{+\infty}
\frac{dt}{t^{n+1}} exp(\frac{1}{2}i(t+\frac{z^2}{t}))
\end{equation}
and summing over all frequencies, it is easy to show that the expression for 
the density matrix can be then rewritten as following,
\begin{equation}
\rho({\bf r},{\bf r'};\beta,mu)=\frac{\delta(r-r')}{2}+
\int d{\bf r^{3P}} 
(\frac{P}{2\pi})^{\frac{3P}{2}} Re (
\{\frac{-2\pi i}{\beta} \sum_{n=0}^{\infty}
(\frac{z_n}{L_P^2})^{\frac{3P}{2}-1}H^{(1)}_{\frac{3P}{2}-1}(z_n)\})
\end{equation}
for fermions
\begin{eqnarray}
\rho({\bf r},{\bf r'};\beta,\mu)=-\frac{\delta(r-r')}{2}-
\int d{\bf r^{3P}}
(\frac{P}{2\pi})^{\frac{3P}{2}} Re (
\{\frac{-2\pi i}{\beta} \sum_{n=1}^{\infty}
(\frac{z_n}{L_P^2})^{\frac{3P}{2}-1}H^{(1)}_{\frac{3P}{2}-1}(z_n) \\
\nonumber +\lim_{w_0\rightarrow +0} \{\frac{-\pi i}{\beta}
(\frac{z_0}{L_P^2})^{\frac{3P}{2}-1}H^{(1)}_{\frac{3P}{2}-1}(z_0)\}) 
\end{eqnarray} 
for bosons, where
\begin{equation}
\begin{array}{lll}
z_n=((K_P^2+i\frac{2w_n}{\beta})L_P^2)^\frac{1}{2} & ;\ \ 
w^{ferm}_n=\pi(2n+1) ;\ \ w^{bos}_n=2\pi n
\end{array}
\end{equation}
and
\begin{eqnarray}
\frac{\hbar^2 K_P^2}{2m}=\mu-\frac{1}{P}
(\frac{V({\bf r})+V({\bf r'})}{2}+\sum_{i=1}^{P-1} V({\bf r^{(i)}}))
\end{eqnarray}  
\begin{eqnarray}
L_P^2=P(({\bf r}-{\bf r^{(1)}})^2+\sum_{i=1}^{P-2}
({\bf r^{(i)}}-{\bf r^{(i+1)}})^2+({\bf r^{(P-1)}}-{\bf r'})^2)
\end{eqnarray}  
In the case of bosons, a term in the summation corresponding to zero 
frequency $w_0$ should be calculated in a limit $w_0 \rightarrow +0$. 

\section{Integral representation}
It is also of interest to show how an integral representation of
the universal functional appears in this formalism. We will outline the
major steps on the example of fermions.
We notice, that for $w_n>0$ equations (\ref{eq1},\ref{eq2}) can be 
written as
\begin{equation}
\frac{1}{iw_n-\beta(H-\mu)}=-i\int_{-\infty}^{0}dt exp(tw_n+it\beta(H-\mu))
\end{equation}
and for $-w_n$
\begin{equation}
\frac{1}{-iw_n-\beta(H-\mu)}=i\int_{0}^{+\infty}dt exp(-tw_n+it\beta(H-\mu))
\end{equation}
Correspondent propagators are
For $w_n>0$
\begin{eqnarray}
<r|\frac{1}{iw_n-\beta(H-\mu)}|r'>=-i\int_{-\infty+i\eta}^{0+i\eta}dt
exp(tw_n) (\frac{imP}{2\pi \hbar t\beta})^{\frac{3P}{2}}
exp(-i(\frac{ml_p^2}{2\hbar t}+\frac{\hbar k_p^2t}{2m}))
\label{eq3}
\end{eqnarray}
and for $-w_n$
\begin{eqnarray}
<r|\frac{1}{-iw_n-\beta(H-\mu)}|r'>=i\int_{0+i\eta}^{+\infty+i\eta}dt
exp(-tw_n) (\frac{imP}{2\pi \hbar t\beta})^{\frac{3P}{2}}
exp(-i(\frac{ml_p^2}{2\hbar t}+\frac{\hbar k_p^2t}{2m}))
\label{eq4}
\end{eqnarray}

Now summation over frequencies can be implemented.For the case of 
fermions 
\begin{equation}
\sum_{n=0}^{+\infty} <r|\frac{1}{iw_n-\beta(H-\mu)}|r'>=
-i\int_{-\infty}^{0} dt (\sum_{n=0}^{+\infty} exp(t\omega_n))
<r|exp(it\beta(H-\mu))|r'>
\end{equation}

\begin{equation}
\sum_{n=-\infty}^{-1} <r|\frac{1}{-iw_n-\beta(H-\mu)}|r'>=
i\int_{0}^{+\infty} dt (\sum_{n=-\infty}^{-1} exp(-t\omega_n))
<r|exp(it\beta(H-\mu))|r'>
\end{equation}

Making an explicit summation over frequencies and substituting the above 
equations into (\ref{eq0}) we get
\begin{equation}
\rho({\bf r},{\bf 
r'};\beta,\mu)=\frac{\delta(r-r')}{2}+\int d{\bf r^{3P}} 
i\int_{-\infty+i\eta}^{+\infty+i\eta} dt W(\beta,t)
(\frac{imP}{2\pi \hbar t\beta})^{\frac{3P}{2}}
exp(-i(\frac{ml_p^2}{2\hbar t}+\frac{\hbar k_p^2t}{2m}))
\end{equation}
where
\begin{equation}
W(\beta,t)=sign(t)\frac{1}{\beta}\frac{exp(-\pi\frac{|t|}{\beta})}
{1-exp(-2\pi\frac{|t|}{\beta})}
\end{equation}
It is straightforward to check that in the limit of $\beta \rightarrow 
\infty$ the above expression coincides with that of Ref.\cite{Harris}.

\section{Scaling of the Universal Functional}
Above we showed that the density matrix for fermions and bosons can be 
expressed via a universal functional $F_{bos, ferm}(\beta,P,L_P^2,K_P^2)$ as
\begin{equation}  
\rho({\bf r},{\bf r'};\beta,\mu)=\pm\frac{\delta(r-r')}{2}\pm\int d{\bf 
r^{3P}} F(\beta,P,L_P^2,K_P^2) 
\end{equation}
From the explicit form of the universal functional derived above one can 
easily derive a scaling law 
\begin{equation}
F(\beta,P,L_P^2,K_P^2)=\beta^{-\frac{3P}{2}} 
F(1,P,L_P^2\beta^{-\frac{1}{2}},K_P^2\beta^{\frac{1}{2}}) 
\end{equation} 

This scaling law can be useful if we consider that the integral
over the universal functional can be rewritten as
\begin{equation}
\int d{\bf r^{3P}} F(\beta,P,L_P^2,K_P^2)=\int dL_P^2 dK_{Po}^2
P(L_P^2,K_{Po}^2) F(\beta,P,L_P^2,\mu+K_{Po}^2)
\end{equation}
where 
\begin{eqnarray}
\frac{\hbar^2 K_{Po}^2}{2m}=-\frac{1}{P}
(\frac{V({\bf r})+V({\bf r'})}{2}+\sum_{i=1}^{P-1} V({\bf r^{(i)}}))  
\end{eqnarray}
and $P(L_P^2,K_{Po}^2)$ is the distribution of pairs of variables
$(L_P^2,K_{Po}^2)$. 
So, in principle, it is enough to calculate the universal functional once 
for say $\beta=1.0$ and the value of the functional at all other 
temperatures can be evaluated using the scaling law. The integration process
can be implemented by first calculating the distribution $P(L_P^2,K_{Po}^2)$
(which actually depends upon ${\bf r}$ and ${\bf r'}$) and then multiplying 
this distribution on the values of the universal functional calculated 
on a 2D grid shifted by $-\mu$ along the $K_{Po}^2$ axis. 

The averages and thermodynamical derivatives of the generic operator 
$U({\bf r},{\bf r'})$ can now be determined from
\begin{equation}
<U({\bf r},{\bf r'})>_{\beta}=\int d{\bf r}d{\bf r'} \rho({\bf r},{\bf 
r'};\beta,\mu) U({\bf r},{\bf r'}) 
\end{equation}
and derivatives from
\begin{equation}
\frac{\partial <U({\bf r},{\bf r'})>_{\beta}}{\partial \beta}=\int d{\bf 
r}d{\bf r'} \frac{\partial \rho({\bf r},{\bf r'};\beta,\mu)}{\partial \beta} 
\end{equation}
\begin{equation}
\frac{\partial <U({\bf r},{\bf r'})>_{\beta}}{\partial \mu}=\int d{\bf
r}d{\bf r'} \frac{\partial \rho({\bf r},{\bf r'};\beta,\mu)}{\partial \mu}
\end{equation}
where the thermodynamic derivatives of the density matrix can be calculated
using the scaling relation for the universal functional giving
\begin{equation}
\frac{\partial \rho({\bf r},{\bf r'};\beta,\mu)}{\partial \mu}=
\pm\frac{1}{\beta^{\frac{3P-1}{2}}} 
\int dL_P^2 dK_{Po}^2 P(L_P^2,K_{Po}^2) 
\frac{\partial 
F(1,P,L_P^2\beta^{-\frac{1}{2}},K_P^2\beta^{\frac{1}{2}})}{\partial 
K_P^2} 
\end{equation}
and
\begin{eqnarray}
\frac{\partial \rho({\bf r},{\bf r'};\beta,\mu)}{\partial \beta}=
\pm\frac{1}{2\beta^{\frac{3P+1}{2}}}     
\int dL_P^2 dK_{Po}^2 P(L_P^2,K_{Po}^2) \\ \nonumber
\{ \frac{\partial 
F(1,P,L_P^2\beta^{-\frac{1}{2}},K_P^2\beta^{\frac{1}{2}})}{\partial K_P^2}-
\frac{1}{\beta}
\frac{\partial 
F(1,P,L_P^2\beta^{-\frac{1}{2}},K_P^2\beta^{\frac{1}{2}})}{\partial L_P^2}-
\frac{3P}{\beta^{\frac{1}{2}}} 
F(1,P,L_P^2\beta^{-\frac{1}{2}},K_P^2\beta^{\frac{1}{2}}) \}
\end{eqnarray} 

\section{A Simple Numerical Study}
To investigate the possible usefulness of this method we have performed
a study of a simple model system consisting of noninteracting fermions in a 
confining potential of the form 
\begin{equation}
V(r)=\frac{10.0}{cosh^2(r)}
\end{equation}
Here we assume to work in units $m=\hbar=1$. This potential allows 5 
discrete levels and analytical solution for the $\Psi$ functions.
\cite{Landau}. The calculations of the density profiles were implemented
as follows. 
First, a universal functional was generated for $\beta=1.0$, $\mu=0.0$,
and $P=20$ (Fig. ~\ref{fig1}). One can see a strong oscillatory nature of the
potential signifying the fermionic nature of the system under consideration.
Next step was to generate the distribution of $P(L_P^2,K_{Po}^2)$ which
is actually the most time consuming part of the calculations. So, to find
$\rho (r,r)$ for each value of $r$ a proper $P(L_P^2,K_{Po}^2)$ distribution
was generated using Monte-Carlo sampling. The final step of our calculations
was in computing the product of the universal potential and the 
$P(L_P^2,K_{Po}^2)$ distribution. By increasing the chemical potential one
moves the $P(L_P^2,K_{Po}^2)$ distribution profile to the right thus 
coupling to more and more oscillations of the universal functional 
henceforth introducing more particles in a system. Thus calculated 
density profiles
for different values of the chemical potential are plotted on Fig. 
~\ref{fig2}. For the sake of comparison density profiles
obtained from the exact solution of Schroedinger equation are also plotted
on Fig. ~\ref{fig3}. Comparing two figures one can notice that even for
such a small number of subdivisions $P=20$ the overall structure
of the density profiles is the same in both cases.

\section{Conclusions}
Grand Canonical formulation of Path Integral Formalism allows an exact 
treatment of fermionic and bosonic systems at any temperature without 
encountering a complicated problem of antisymmetrization or 
symmetrization of coordinate eigen states. In this formulation density 
matrix is expressed as an integral over the universal functional 
$F_{bos,ferm}(\beta,P,L_P^2,K_P^2)$ while particular properties of a 
system come through the variables $L_P^2,K_P^2$.  
Computationally convenient representation of the universal functional is 
possible in terms of Hankel functions of the first kind so one can, in
principle, precalculate the $F_{bos,ferm}(\beta,P,L_P^2,K_P^2)$ and the 
density itself will be determined by a distribution of the non-universal 
variables $L_P^2,K_P^2$.

\begin{figure}
\caption{$F_{ferm}(\beta=1.0,P=20,L_P^2,K_{Po}^2)$. One can easily notice 
strong oscillations due to the nature of fermionic statistics.
\label{fig1}}
\end{figure}

\begin{figure}
\caption{
Density profiles for different values of the chemical potential calculated
using Grand Canonical Path Integral Method.
\label{fig2}}
\end{figure}
 
\begin{figure}
\caption{
Density profiles for different values of the chemical potential calculated using 
the exact solution of Schroedinger equation.
\label{fig3}}
\end{figure}

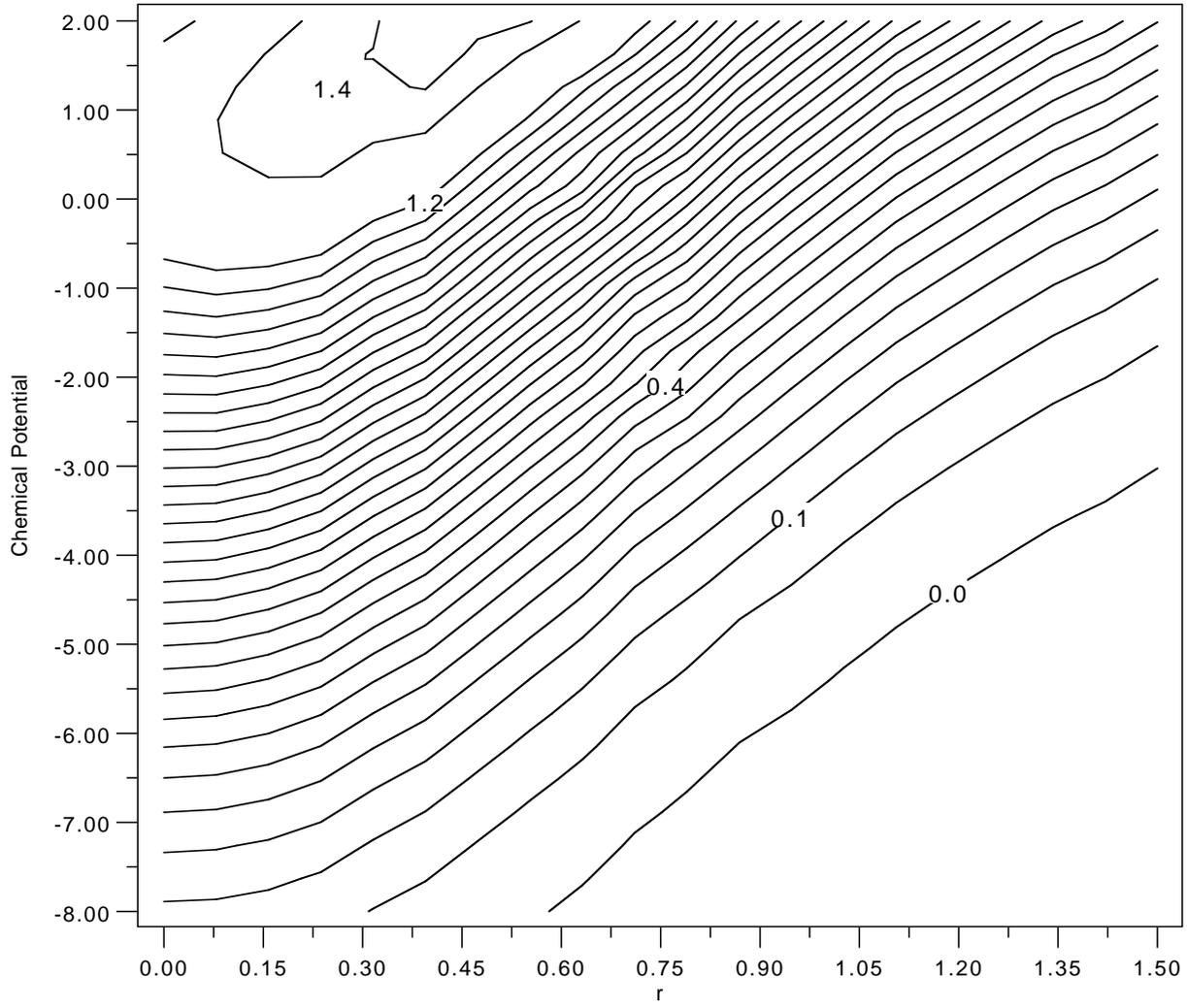
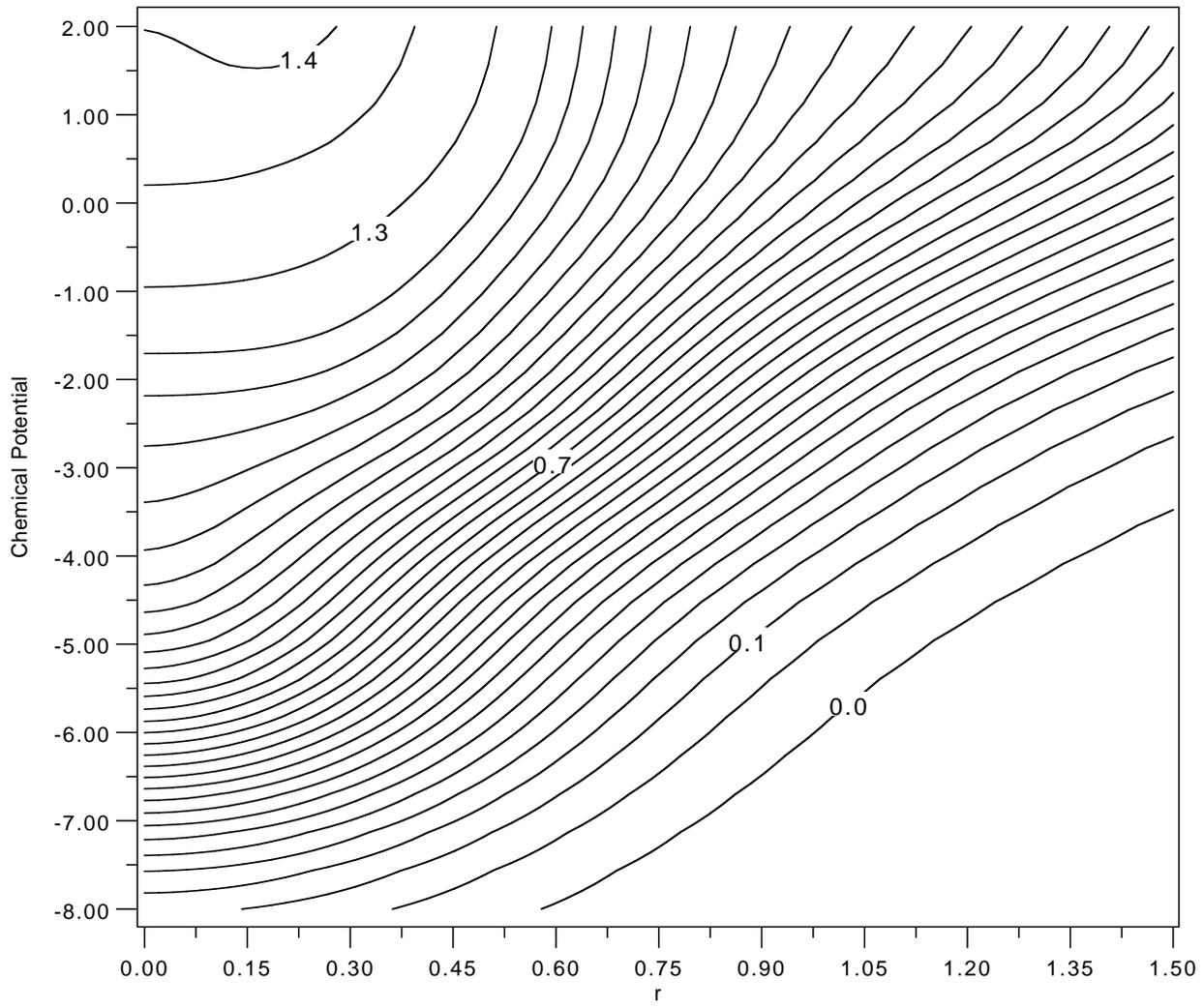
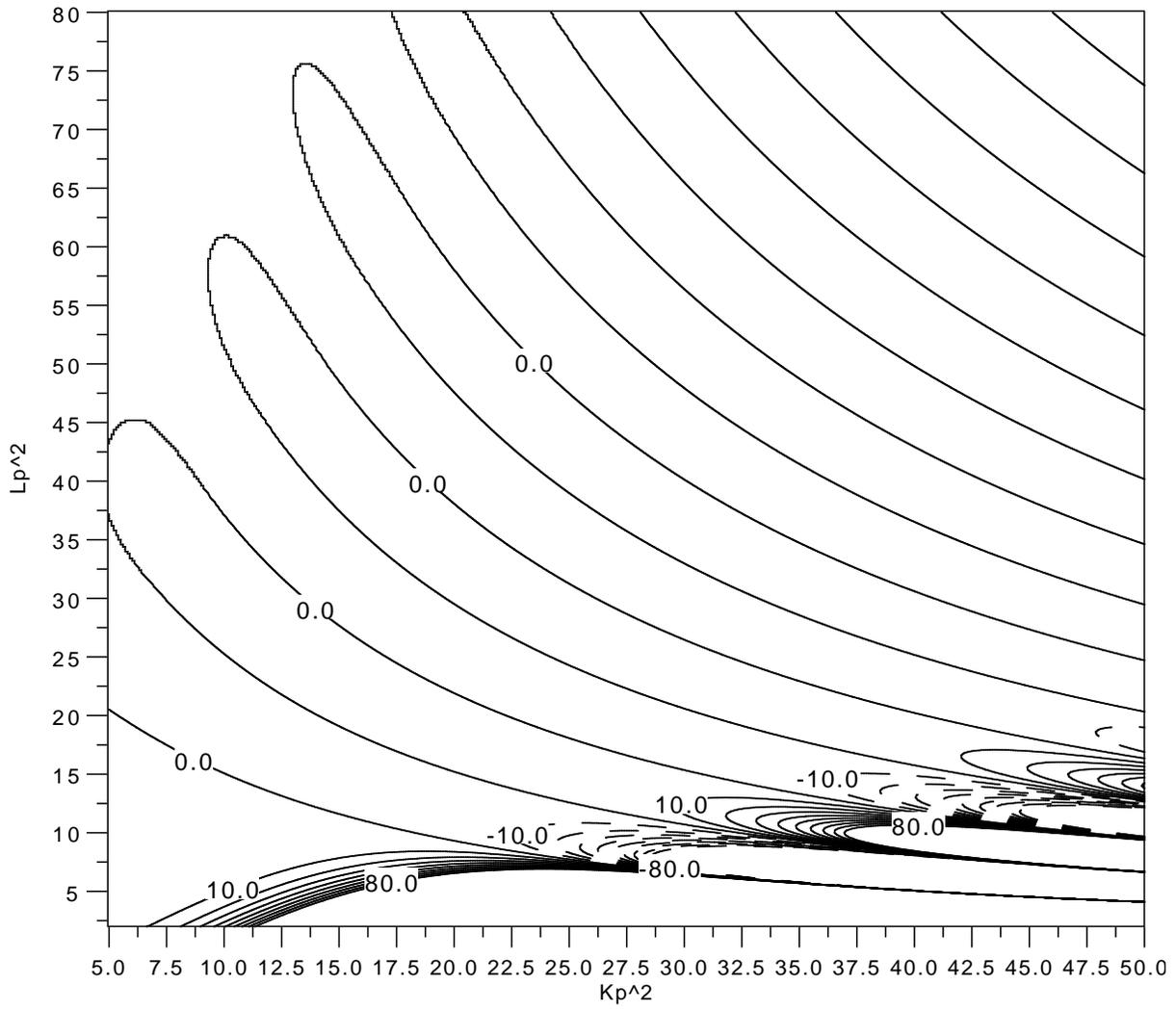
\end{document}